\def\breakon{\end{multicols}\widetext\vspace{-.6cm}
\noindent\rule{.49\linewidth}{.3mm}\rule{.3mm}{.5cm}\vspace{0.0cm}}
\def\breakoff{\vspace{-0.45cm}
\noindent
\rule{.50\linewidth}{.0mm}\rule[-.47cm]{.3mm}{.5cm}\rule{.49\linewidth}{.3mm}
\vspace{-0.25cm}
\begin{multicols}{2}   }
\documentstyle[aps,epsf,multicol]{revtex}
\begin{document}

%%%%%%%%%%%%%%%%%%%%%%%%%%%%%%%%%%%%%%%%%%%%%%%%%%%%%%%%%%
%% multicol commands
%%%%%%%%%%%%%%%%%%%%%%%%%%%%%%%%%%%%%%%%%%%%%%%%%%%%%%%%%%
% This is needed so that figure captions come out right 
% when using the "multicols" setting.
\makeatletter
\renewenvironment{table}
  {\let\@capwidth\linewidth\def\@captype{table}}
  {}

\renewenvironment{figure}
  {\let\@capwidth\linewidth\def\@captype{figure}}
  {}
\makeatother
%%
%%%%%%%%%%%%%%%%%%%%%%%%%%%%%%%%%%%%%%%%%%%%%%%%%%%%%%%%%%

\title{Kondo Effect in Carbon Nanotube Single-Electron Transistors}
\author{Eugene H. Kim~$^{1}$, Germ\`{a}n Sierra~$^{2}$, 
        and C. Kallin~$^{1}$ }
\address{$^1$ Department of Physics and Astronomy, 
         McMaster University, Hamilton, Ontario, Canada L8S-4M1  \\
         $^2$ Instituto de Matem\`{a}ticas y F\`{i}sica Fundamental,
         C.S.I.C., 28006 Madrid, Spain }
\maketitle

\begin{abstract}
Recently, Coulomb blockade physics was observed at room temperature
in a carbon nanotube single-electron transistor (H.~W.~Ch. Postma, 
{\it et. al.}, Science {\bf 293}, 76 (2001)).  In this work, we 
suggest that these devices may be promising for studying the Kondo 
effect.  In particular, they could allow for a detailed investigation
of the 2-channel Kondo fixed point.  Moreover, fabricating a similar 
device in a short nanotube could be promising for studying the effect 
of a magnetic impurity in an ultrasmall metallic grain.  Experimental
signatures of the Kondo effect in these systems is discussed.
\end{abstract}

%%%%%%%%%%%%%%%%%%%%%%%%%%%%%%%%%%%%%%%%%%%%%%%%%%%%%%%%%%%%%%%%%%%%%%%%%%%
%%%%%%%%%%%%%%%%%%%%%%%%%%%%%%%%%%%%%%%%%%%%%%%%%%%%%%%%%%%%%%%%%%%%%%%%%%%

\vspace{.0in}
\begin{multicols}{2}

Recently, carbon nanotubes have been the source of an enormous
amount of activity.\cite{reviews}  The remarkable control with
which these materials can be fabricated and manipulated makes
carbon nanotubes an ideal system for studying the electronic
properties of one-dimensional conductors.  Moreover, these 
materials are extremely durable, and relatively inexpensive 
to make.  Therefore, besides fundamental science, these systems 
are promising for commercial applications.

In recent work,\cite{postma} a single-electron transistor (SET) 
was fabricated by introducing two buckles in series in a long 
single-wall carbon nanotube.  The two buckles define a small 
island ({\it i.e.} a ``quantum dot'') within the nanotube.  
(See Fig. 1 of Ref.~\ref{postma}).  
Using this device, the authors of Ref.~\ref{postma} observed 
Coulomb blockade physics at room temperature.  Moreover, they 
found that the conductance had a power-law temperature dependence, 
consistent with a Luttinger liquid model for the leads.  In this 
work, we suggest that this device could be promising for studying 
the Kondo effect.

The Kondo effect in Coulomb blockade systems has received a 
considerable amount of attention over the last few years.\cite{leo}  
However, in most of these studies, the leads were described by 
non-interacting electron gases.  The case where the leads 
themselves are interacting liquids has only recently received 
attention, and has been shown to exhibit rich behavior driven 
by these interactions.\cite{simon,gene}  Carbon nanotube SETs 
could provide a controlled environment for studying the Kondo 
effect in systems with interacting leads. 
It should be noted that carbon nanotubes have also been 
shown to display interesting mesoscopic effects, characteristic
of nanoscale conductors.\cite{reviews}
Recently, it has been shown that interesting physics would 
arise if a magnetic impurity were placed in an ultrasmall 
metallic grain, due to the finite level spacing of the 
grain.\cite{vondelft}  (In Ref.~\ref{vondelft}, this system 
was dubbed the {\sl Kondo box}.)  A device similar to the one 
used in Ref.~\ref{postma} could provide a controllable realization 
of the Kondo box.

%%%%%%%%%%%%%%%%%%%%%%%%%%%%%%%%%%%%%%%%%%%%%%%%%%%%%%%%%%%%%%%%%%%%%%%%%%%
%%%%%%%%%%%%%%%%%%%%%%%%%%%%%%%%%%%%%%%%%%%%%%%%%%%%%%%%%%%%%%%%%%%%%%%%%%%

We begin our discussion by recalling the band structure of 
carbon nanotubes.  These materials consist of a sheet of 
graphite rolled into a cylinder.  A single sheet of graphite 
consists of carbon atoms arranged on the sites of a honeycomb 
lattice.  The band structure is well described by a tight-binding 
model with one orbital per lattice site.  To form a nanotube, 
the sheet of graphite is rolled into a cylinder.  Doing this 
quantizes the crystal momentum, $q_y$, transverse to the axis 
of the cylinder.  Interestingly, two (one-dimensional) bands 
of gapless excitations exist at $q_y = 0$.
The low energy physics is determined by the two bands of gapless 
excitations (labeled as band-$c$ and band-$d$), which disperse 
with the same velocity.  

%%%%%%%%%%%%%%%%%%%%%%%%%%%%%%%%%%%%%%%%%%%%%%%%%%%%%%%%%%%%%%%%%%%%%%%%%%%
%%%%%%%%%%%%%%%%%%%%%%%%%%%%%%%%%%%%%%%%%%%%%%%%%%%%%%%%%%%%%%%%%%%%%%%%%%%

With regards to interactions, interbranch ({\it i.e.} backscattering)
interactions are weak.  These interactions are determined by the short 
range part of the Coulomb interaction.  However, the probability of 
two electrons being near each other is suppressed in the two low energy
bands, since these bands have $q_y = 0$ and hence are extended around 
the circumference of the tube.
For isolated single-wall nanotubes, however, the Coulomb interaction 
is unscreened.  Therefore, to describe the system in Ref.~\ref{postma}, 
one must take into account the long range nature of the Coulomb 
interaction.

A considerable amount is known about the two-band model of interacting
electrons.\cite{lin}  In the undoped case interactions drive the system 
to a Mott-insulating state, with a gap to both spin and charge excitations.  
When doped with holes, the spin gap remains and the holes form pairs.  
In nanotubes, the (backscattering) interactions which drive these 
instabilities are weak.  Hence, these effects will only be observable 
at very low temperatures/energies.  Above the spin gap and pairing 
energy scale, the system behaves as a Luttinger liquid.\cite{kane}

The spin gap introduces complications for the Kondo effect.  
However, since the spin gap in carbon nanotubes is small, it
can be overcome by a modest magnetic field.  
The main effect of a magnetic field is to shift the 
bands of the ``spin-up'' and ``spin-down'' electrons.
Because of this, the processes which cause the spin gap suffer a 
momentum mismatch and become irrelevant.  The only processes which 
survive are triplet pairing interactions.  
The results of Ref.~\ref{cabra} suggest that the triplet pairing
interactions are marginally relevant, but the energy scale at 
which their effects are visible is unattainably low.  Therefore, 
we will ignore them.  Although the competition of the spin gap and 
the Kondo effect is an interesting issue, in this work we will 
focus on the case where there are always low energy spin excitations 
present.

%%%%%%%%%%%%%%%%%%%%%%%%%%%%%%%%%%%%%%%%%%%%%%%%%%%%%%%%%%%%%%%%%%%%%%%%%%%
%%%%%%%%%%%%%%%%%%%%%%%%%%%%%%%%%%%%%%%%%%%%%%%%%%%%%%%%%%%%%%%%%%%%%%%%%%%

In Ref.~\ref{postma}, an SET was fabricated by creating
a small island within a long single-wall carbon nanotube.
Being interested in the low energy properties of the system, 
we focus on the uppermost level of the island and model it 
as an Anderson impurity.  
The Hamiltonian, including the coupling to the leads, is 
\begin{eqnarray}
 & &  H_{\rm island} = \varepsilon_0 \sum_s n^f_s 
  + U_0~ n^f_{\uparrow} n^f_{\downarrow} 
   - \frac{h_0}{2} \left(n^f_{\uparrow} - n^f_{\downarrow} \right)  
  \label{anderson}  \\ & & \hspace{.2in}
  - \sum_{\stackrel{\lambda = c,d}{s=\uparrow,\downarrow}}
  \left(
  t^{\phantom \dagger}_{1\lambda} \psi^{\dagger}_{1,\lambda,s}(0) 
  +~ t^{\phantom \dagger}_{2\lambda} \psi^{\dagger}_{2,\lambda,s}(0) 
  \right) f^{\phantom \dagger}_s + h.c.  \, ,  \nonumber  
\end{eqnarray}
where $\psi_{i,\lambda,s}$ destroys an electron with spin-$s$ 
in lead-$i$ ($i=1,2$) and band-$\lambda$ ($\lambda=c,d$); 
$f^{\phantom \dagger}_s$ destroys an electron with spin-s on 
the island; $n^f_{s} = f^{\dagger}_s f^{\phantom \dagger}_s$;
$\varepsilon_0$ is the energy level of the island, which can 
be controlled by a gate voltage; $U_0$ is the charging energy; 
$h_0$ is the magnetic field; $t_{i\lambda}$ is the matrix 
element for an electron to tunnel to the island from 
band-$\lambda$ in lead-$i$.
It is useful to introduce {\sl bonding} and {\sl antibonding} 
combinations
\begin{eqnarray}
 \psi_{i,b,s} & = & \left(t_{ic}~\psi_{i,c,s} 
 + t_{id}~\psi_{i,d,s} \right)/\sqrt{N_i} \, ,
  \nonumber \\
 \psi_{i,a,s} & = & \left(t_{id}~\psi_{i,c,s} 
 - t_{ic}~\psi_{i,d,s} \right)/\sqrt{N_i} \, ,
\label{newfermions}
\end{eqnarray}
with $N_i^{\phantom 2} = t_{ic}^2 + t_{id}^2$.  In terms of these
operators, we see that only the bonding combinations couple to the 
island.
Being interested in the Kondo regime, we integrate out charge 
fluctuations on the island.  Working to second order in perturbation 
theory,\cite{schreiffer} we arrive at the effective Hamiltonian
\begin{eqnarray}
 & H_{\rm int} & ~ =~ {\bf \tau} \cdot \frac{\sigma_{s,s'}}{2}
   \left( J_1^{\phantom \dagger} 
   \psi^{\dagger}_{1,b,s}(0) \psi^{\phantom \dagger}_{1,b,s'}(0)
  + 1 \rightarrow 2 \right) \nonumber \\   
 & + & J_{12}^{\phantom \dagger}{\bf \tau} \cdot \frac{\sigma_{s,s'}}{2}
   \left(\psi^{\dagger}_{1,b,s}(0) \psi^{\phantom \dagger}_{2,b,s'}(0) 
 + h.c. \right) - h_0~ \tau_z \, , 
\label{schreifferwolff}
\end{eqnarray} 
where $\tau$ is the spin operator for the electron on the island,
and the values of the couplings ($J_i$ and $J_{12}$)
can be found in {\it e.g.} Ref.~\ref{simon}.  It is important to
note, however, that $J_i > 0$ and $J_{12} > 0$.
It should also be noted that in Eq.~\ref{schreifferwolff} we have 
not displayed the potential scattering terms\cite{schreiffer} which 
were generated.  For the system considered in this work, these terms 
have a very small effect and can be ignored.\cite{gene}

The dynamics of the leads is described by the Hamiltonian 
$H_{\rm leads} = H_{\rm lead-1} + H_{\rm lead-2}$, where 
$H_{{\rm lead}-i} = H_{i}^0 + H_{i}^1$ is the Hamiltonian 
for lead-$i$ with\cite{kane}
\begin{eqnarray}
  H_{i}^0 & = & - i v_F \sum_{\lambda , s} \int_{-l}^0 
   \hspace{-0.075in} dx \left( \psi^{\dagger}_{R,i,\lambda,s}\partial_x 
    \psi^{\phantom \dagger}_{R,i,\lambda,s} - R \rightarrow L
   \right)   
  \label{lead}  \\  
  H_{i}^1 & = & U \int_{-l}^0 \hspace{-.075in} dx \left(\sum_{\lambda,s} 
   \psi^{\dagger}_{R,i,\lambda,s}\psi^{\phantom \dagger}_{R,i,\lambda,s} 
 + \psi^{\dagger}_{L,i,\lambda,s}\psi^{\phantom \dagger}_{L,i,\lambda,s}
   \right)^2  \, .  \nonumber  
\end{eqnarray}
In the above equation, $\psi_{R,i,\lambda,s}$ ($\psi_{L,i,\lambda,s}$)
is the right (left) moving component of $\psi_{i,\lambda,s}$. 
Furthermore, we have followed Ref.~\ref{kane} and taken the Coulomb 
interaction to be screened beyond some long distance; $U$ is the 
effective strength of this interaction.
In the previous paragraph, we saw that only the bonding combination 
of the fermion fields (Eq.~\ref{newfermions}) couples to the impurity.  
Fortunately, we can express the Hamiltonian of the leads in terms of 
the bonding and antibonding operators as well.  In terms of these 
operators, the Hamiltonian has the same form as Eq.~\ref{lead}, except
the labels $c$ and $d$ are replaced everywhere by $b$ and $a$.

In what follows, we will make extensive use of the boson 
representation.  To do so, the electron operator is written as 
$\psi_{R/L,i,\lambda,s} \sim 
   e^{\pm i \sqrt{4\pi} \phi_{R/L,i,\lambda,s} }$
where the chiral fields, $\phi_{R,i,\lambda,s}$ and 
$\phi_{L,i,\lambda,s}$, are related to the usual Bose field 
$\phi_{i,\lambda,s}$ and its dual field $\theta_{i,\lambda,s}$ by 
$\phi_{i,\lambda,s} = \phi_{R,i,\lambda,s} + \phi_{L,i,\lambda,s}$ and 
$\theta_{i,\lambda,s} = \phi_{R,i,\lambda,s} - \phi_{L,i,\lambda,s}$.  
It will also prove useful to form {\sl charge} and {\sl spin} fields
$\phi_{i,\lambda,\rho/\sigma} = \left(\phi_{i,\lambda,\uparrow} 
 \pm \phi_{i,\lambda,\downarrow} \right)/\sqrt{2}$,
and then form the combinations
$\phi_{i,\rho^{\pm}} = \left(\phi_{i,b,\rho} \pm 
 \phi_{i,a,\rho} \right)/\sqrt{2}$ 
describing {\sl total} and {\sl relative} 
charge fluctuations in lead-$i$.
In terms of these variables, the Hamiltonian for lead-$i$ is
\begin{eqnarray}
 H_{{\rm lead}-i} & = & \frac{v_{\rho^+}}{2} \int_{-l}^0 dx~  
  K_{\rho^+} \left(\partial_x \theta_{i,\rho^+} \right)^2 +
  \frac{1}{K_{\rho^+}} \left(\partial_x \phi_{i,\rho^+} \right)^2
 \nonumber \\
 & + & \frac{v_F}{2} \int_{-l}^0 dx~
   \left(\partial_x \theta_{i,\rho^-} \right)^2
 + \left(\partial_x \phi_{i,\rho^-} \right)^2   \\
 & + & \frac{v_F}{2} \sum_{\lambda=b,a} \int_{-l}^0 dx~ 
    \left(\partial_x \theta_{i,\lambda,\sigma} \right)^2  
  + \left(\partial_x \phi_{i,\lambda,\sigma} \right)^2  
  \, ,   \nonumber  
\end{eqnarray}
where $K_{\rho^+} = 1/\sqrt{1 + 8U/(\pi v_F)}$ and 
$v_{\rho^+} = v_F/K_{\rho^+}$.  Experimentally, 
it has been found that $0.19 \leq K_{\rho^+} \leq 0.26$ for 
single-wall carbon nanotubes.\cite{postma}
Finally, to analyze the physics it will prove useful to unfold 
the system, and work solely in terms of right moving fields.\cite{eggert}

%%%%%%%%%%%%%%%%%%%%%%%%%%%%%%%%%%%%%%%%%%%%%%%%%%%%%%%%%%%%%%%%%%%%%%%%%%%
%%%%%%%%%%%%%%%%%%%%%%%%%%%%%%%%%%%%%%%%%%%%%%%%%%%%%%%%%%%%%%%%%%%%%%%%%%%

We begin our discussion of the Kondo effect by considering the
case of semi-infinite leads: $l \rightarrow \infty$.
Near the ultraviolet fixed point, we can compute the conductance
using the golden rule.  We find $G \sim T^{\alpha}$, where 
$\alpha = (1/2)(1/K_{\rho^+} - 1)$, in agreement with what was 
reported in Ref.~\ref{postma}.  The behavior of the system at 
lower energies can be deduced by a renormalization group (RG) 
analysis. 
To second order in the couplings,\cite{cardy} the RG equations 
for the parameters are
\begin{eqnarray}
 \frac{d\lambda_{+}}{dl} & = & \lambda_{+}^2 + \lambda_{-}^2 + g^2  
 \ \ \ , \ \ \
 \frac{d\lambda_{-}}{dl} = 2 \lambda_{+} \lambda_{-}  \, ,
 \nonumber \\
 \frac{dg}{dl} & = &  
    \frac{1}{4}\left(1 - \frac{1}{K_{\rho^+}} \right) g 
    + 2 g \lambda_{+} 
 \ \ \ , \ \ \
 \frac{d \lambda_h}{dl} = \lambda_h  \, , 
\label{ultravioletRG}
\end{eqnarray}
where $\lambda_{+} \sim (J_1 + J_2)$, 
$\lambda_{-} \sim (J_1 - J_2)$,  $g \sim J_{12}$, and 
$\lambda_h \sim h_0$.
A few words are in order about the RG equations.  Let us 
first consider $J_1 = J_2$, so that $\lambda_{-} = 0$.  At 
the ultraviolet fixed point, the $J_1$ and $J_2$ terms are 
marginally relevant.  On the other hand, the $J_{12}$ term 
is irrelevant for repulsive interactions ($K_{\rho^+} < 1$).  
Hence, $g$ will initially decrease under the RG.  For the 
values of $K_{\rho^+}$ relevant to this system, $\lambda_{+}$ 
will have grown to ${\cal O}(1)$ while $g \ll 1$.\cite{gene}  
If $g=0$, we would have a 2-channel Kondo model, which is 
known to have a nontrivial ${\cal O}(1)$ fixed point.  
Therefore, for $J_1 = J_2$ the low energy physics will be 
governed by the 2-channel Kondo fixed point with $g$ (and 
$\lambda_h$) as perturbations.  
Now let us consider $J_1 \neq J_2$.  From Eq.~\ref{ultravioletRG}, 
$\lambda_{-}$ will grow under the RG.  If $J_1$ and $J_2$ are 
considerably different (for concreteness, consider $J_1 > J_2$), 
the system will flow to the 1-channel Kondo fixed point, where 
the electron on the island forms a singlet with the electrons in 
lead-$1$.\cite{blandin}  However, for $J_1 \approx J_2$, 
$\lambda_{-}$ will grow slowly, so that the system flows close to 
the 2-channel Kondo fixed point.  In this case, it is appropriate 
to consider the behavior near the 2-channel Kondo fixed point with 
$g$ and $\lambda_{-}$ (and $\lambda_h$) as perturbations.
Since the device we are considering is made by introducing buckles
in a carbon nanotube, it will probably be difficult to achieve
$J_1 = J_2$.  However, as we feel the possibility of observing 
2-channel Kondo physics is one of the most interesting features 
of this system, in what follows we will focus on the case
$J_1 \approx J_2$.      
Finally, it should be noted that the magnetic field is a relevant 
perturbation.  Therefore, we must consider very small fields, so 
as not to completely wipe out the Kondo physics described above. 

%%%%%%%%%%%%%%%%%%%%%%%%%%%%%%%%%%%%%%%%%%%%%%%%%%%%%%%%%%%%%%%%%%%%%%%%%%%
%%%%%%%%%%%%%%%%%%%%%%%%%%%%%%%%%%%%%%%%%%%%%%%%%%%%%%%%%%%%%%%%%%%%%%%%%%%

To analyze the physics near the 2-channel Kondo fixed point,
we follow Ref.~\ref{schiller} and form combinations of the 
fields in the two leads: $\phi_{R,c}$, $\phi_{R,sp}$, $\phi_{R,f}$, 
and $\phi_{R,sf}$.  
Then, we perform the unitary transformation,
$U = \exp \left( i \sqrt{4\pi}~ \tau^z \phi_{R,sp}(0) \right)$,
which ties a spin-1/2 from the leads to the island.
% \cite{emery}
Finally, we introduce new fermion fields, $d \sim S^-$ and
$X \sim e^{i\sqrt{4\pi}\phi_{R,sf}}$.
Upon performing these transformations, $H_{\rm int}$ becomes
\begin{eqnarray}
 H_{\rm int} & = & v_F \lambda_{+}' \left( d^{\dagger} - d~ \right) 
   \left( X^{\dagger}(0) + X(0) \right)     
 \label{toulouse} \\
 & & \hspace{-.475in} + v_F \lambda_{-}' \left( d^{\dagger} + d~ \right) 
   \left( X^{\dagger}(0) - X(0) \right)  
 - v_F \lambda_h' \left( d^{\dagger} d^{\phantom \dagger} - 1/2 \right)
 \nonumber \\
 & & \hspace{-.4in} + v_F g' \left( d^{\dagger} + d~ \right)
  \left( e^{-i\sqrt{4\pi}\phi_{R,f}(0)} 
  - e^{i\sqrt{4\pi}\phi_{R,f}(0)} \right) 
% \nonumber \\
 \, , \nonumber   
\end{eqnarray}
where $\lambda_{+}'$, $\lambda_{-}'$, $g'$, and $\lambda_h'$ are 
the renormalized values of the couplings.  Note that in 
Eq.~\ref{toulouse}, we have displayed only the most relevant 
operators.
A few words are in order about Eq.~\ref{toulouse}.  To begin with,
the $\lambda_{+}'$ term sets the 2-channel Kondo energy scale; the 
$g'$, $\lambda_{-}'$, and $\lambda_h'$ terms are perturbations about 
the 2-channel Kondo fixed point.  The $g'$ term has dimension 
$(1 + 1/K_{\rho^+})/4$, and is relevant for $K_{\rho^+} > 1/3$.
Hence, this term is irrelevant for the system we are considering.  
Both the $\lambda_{-}'$ and $\lambda_h'$ terms have dimension 1/2 
and are relevant.  If these terms are absent, the zero temperature
fixed point would be the 2-channel Kondo fixed point.  However,
nonzero $\lambda_{-}'$ and/or $\lambda_h'$ drives the system away
from the two-channel Kondo fixed point.  $\lambda_{-}'$ drives the 
system to the 1-channel Kondo fixed point, where the electron on 
the island forms a singlet with the electrons in the lead with the 
larger exchange coupling.\cite{blandin,ludwig}  The $\lambda_h'$ 
term drives the system to a fixed point where the electron on the 
island is spin polarized; spin-flip processes are energetically 
costly, and the electron on the island behaves as a potential 
scatterer.\cite{ludwig}  The energy scale at which 2-channel Kondo 
behavior will no longer be observable is determined by the values 
of $\lambda_{-}'$ and $\lambda_h'$.

\begin{figure}
\epsfxsize=3.0in
\centerline{\epsfbox{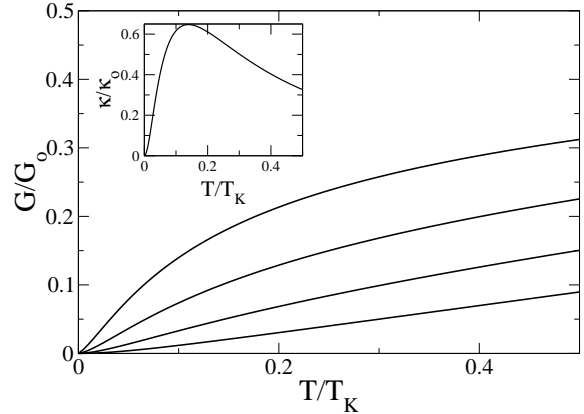} }
\vspace{.2in}
\caption{$G/G_0$ vs. $T/T_K$  near the 2-channel Kondo fixed 
point.  $K_{\rho} = 0.29, 0.26, 0.23, 0.2$ in order from the 
top to the bottom curve.
Inset: $\kappa / \kappa_0$ vs. $T/T_K$ near the 2-channel Kondo
fixed point.
In both plots, the parameters $\Gamma_{-}$ and $\Gamma_h$ were 
taken to be $\Gamma_{-} = 0.07$ and $\Gamma_h = 0.1$.  }
\label{fig:plot2ck}
\end{figure}

Signatures of the 2-channel Kondo fixed point can be observed 
in conductance measurements.  Using the golden rule, we find
\begin{eqnarray}
 & & G/G_0~ = \frac{1}{\Gamma(\beta)} 
  \left(\frac{T}{T_K}\right)^{\beta - 2}
  \int \frac{dx}{2\pi}~  {\rm sech}\left(\frac{x~T_K}{2T}\right)  
 \label{G2ck}  \\  & \times &
   \left| \Gamma\left(\frac{\beta}{2} 
     + i \frac{x~ T_K}{2\pi T}\right) \right|^2  \hspace{-0.06in} 
   \frac{\Gamma_{-} (1 + x^2) + \Gamma_h}
     {(x^2 - \Gamma_h - \Gamma_{-})^2 + x^2 (1 + \Gamma_{-})^2} \, .  
\nonumber 
\end{eqnarray} 
In Eq.~\ref{G2ck}, $T_K = E_0 \exp(-1/\lambda_+)$, where $E_0$ is 
a high-energy cut-off; $\beta = (1/2)(1 + 1/K_{\rho^+})$; 
$G_0 = (2e^2/h)(g')^2/(2\pi)$; $\Gamma_{-} \sim (\lambda_{-}')^2$; 
$\Gamma_h \sim (\lambda_h')^2$.  $G/G_0$ vs. $T/T_K$ is plotted in  
Fig.~\ref{fig:plot2ck} for several values of $K_{\rho^+}$.  To begin 
with, notice that the conductance decreases as the temperature is
decreased.  This should be contrasted with the case of non-interacting 
leads, where the Kondo effect leads to perfect conductance at low 
temperatures.\cite{leo}  This behavior is due to the interactions 
in the leads.  From Eq.~\ref{G2ck}, it follows that $G \sim T^{\beta - 2}$ 
for $\Gamma_h, \Gamma_{-} \ll T \ll T_K$.  This temperature dependence 
is a property of the 2-channel Kondo fixed point.  However, for 
$T < \Gamma_{-}$ and/or $T < \Gamma_h$, the system is far from the
2-channel Kondo fixed point, and the temperature dependence is
modified from its 2-channel Kondo behavior.

\vspace{.1in}

Besides the (charge) conductance, 2-channel Kondo physics can also 
be observed in thermal conductance measurements.  An interesting 
property of the 2-channel Kondo fixed point is that it has perfect 
spin conductance.\cite{gene}  Though the spin conductance is difficult 
to measure, this will manifest itself in the thermal conductance 
--- as charge transport is suppressed, the thermal conductance will 
be dominated by spin.  Computing the thermal conductance\cite{sivan} 
due to spin, we find 
\begin{eqnarray}
  \kappa / \kappa_0 & = & \left(\frac{3}{4\pi^2}\right) 
  \left(\frac{T_K}{T}\right)^3 \int dx~ 
  {\rm sech}^2\left(\frac{x~T_K}{2T}\right) 
 \label{kappa2ck}  \\  & & \hspace{.15in} \times
 \frac{x^4 (1- \Gamma_{-})^2}
 { (x^2 - \Gamma_h - \Gamma_{-})^2 + x^2 (1 + \Gamma_{-})^2}  \, ,   
 \nonumber
\end{eqnarray}
where $\kappa_0 = (\pi^2/3) T/ h$ is the value for perfect thermal
conductance.  $\kappa/\kappa_0$ vs. $T/T_K$ is shown in the inset 
of Fig.~\ref{fig:plot2ck}.  From Eq.~\ref{kappa2ck},  
$\kappa \rightarrow \kappa_0$ for $\Gamma_h,\Gamma_{-} \rightarrow 0$ 
(for $T \ll T_K$).  This is due to the perfect spin conductance of 
the 2-channel Kondo fixed point.  However, $\Gamma_{-} \neq 0$ 
and/or $\Gamma_h \neq 0$ drives the system away from the 2-channel 
Kondo fixed point and destroys the perfect spin conductance.

\begin{figure}
\centerline{\epsfxsize=2.75in \epsfbox{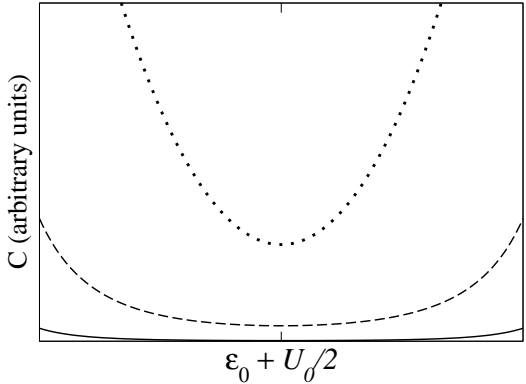} } 
\vspace{.2in}
\caption{Differential capacitance vs. gate voltage ---  
dotted line: island coupled to semi-infinite leads; dashed line:
Kondo box with $N$ = even; solid line: Kondo box with $N$ = odd.}
\label{fig:capacitance}
\end{figure}

Another way to probe the Kondo physics is by measuring the 
differential capacitance as a function of gate voltage.\cite{ashoori}  
At $T=0$, $C = \partial^2 E_G / \partial V_G^2$, where $E_G$ is 
the ground-state energy and $V_G$ is the gate voltage coupled to 
the island.  Furthermore, we expect 
$\varepsilon_0 = \eta V_G + {\rm const}$, where $\eta$ is a constant.  
For $\Gamma_{-}, \Gamma_h \ll T_K$, the contribution to the ground 
state energy due to the Kondo effect is
$\delta E_G \approx (\ln (c_0)/2\pi)~ T_K $, where $c_0$ is a constant
of order unity.
Differentiating, we find 
\begin{equation}
 C \sim \varepsilon_0^2 
  \exp\left(\frac{\pi \varepsilon_0(\varepsilon_0 + U_0)}
 {U_0\Gamma_0}\right) \, ,
\end{equation}
where $\Gamma_0 = 2(t_{1c}^2 + t_{1d}^2 + t_{2c}^2 + t_{2d}^2)/v_F$.
This strongly varying function of gate voltage is due to 
the Kondo effect.  The differential capacitance vs. gate 
voltage is plotted in Fig.~\ref{fig:capacitance}.  

\vspace{.1in}
\vspace{.1in}

%%%%%%%%%%%%%%%%%%%%%%%%%%%%%%%%%%%%%%%%%%%%%%%%%%%%%%%%%%%%%%%%%%%%%%%%%%%
%%%%%%%%%%%%%%%%%%%%%%%%%%%%%%%%%%%%%%%%%%%%%%%%%%%%%%%%%%%%%%%%%%%%%%%%%%%

\begin{figure}
\centerline{\epsfxsize=1.85in \epsfbox{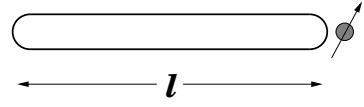} } 
\vspace{.2in}
\caption{ Schematic of the Kondo box configuration:
a small island coupled to a larger nanotube ``nanoparticle''. }
\label{fig:finite}
\end{figure}

Now we consider the Kondo effect in a short nanotube --- a Kondo 
box.  More specifically, we consider a short carbon nanotube with 
a buckle introduced near one of the ends.  This buckle defines a 
small island, which is connected to a larger nanotube ``nanoparticle'' 
of length $l$.  (See Fig.~\ref{fig:finite}.)  For this configuration, 
$t_{2c} = t_{2d} = 0$ in Eq.~\ref{anderson}.  Also, to simplify 
things let us consider $h_0 = 0$.  Then, only $J_1 \neq 0$ while 
$J_2 = 0$ and $J_{12} = 0$ in Eq.~\ref{schreifferwolff}. 
The Kondo effect in this system can be probed by measuring the 
differential capacitance as a function of gate voltage.  Here, 
we find that the results strongly depend on the total number of 
particles in the system, $N$.  ($N$ = number of electrons in the 
nanoparticle $+$ electron on the island.)  Calculating the shift 
in the ground state energy, we find
\begin{eqnarray}
 \delta E_G & = & - \frac{3}{4} \frac{\Delta}
  {\ln\left(\frac{\Delta}{T_K}\right)}  
  \ \ \ \ \ \ \ {\rm N = even}   \, , \nonumber \\
 \delta E_G & = & - \frac{2.7}{160} \frac{\Delta}
  {\ln^2\left(\frac{\Delta}{T_K}\right)} 
  \ \ \ \ {\rm N = odd}  \, ,
\label{boxenergy}
\end{eqnarray}
where $\Delta = v_F \pi / l$ is the level spacing, and we 
are assuming $T_K \ll \Delta$.  Notice that $\delta E_G$ 
is significantly greater for $N$=even as compared with 
$N$=odd.  This occurs because for $N$=even, the ground 
state of the nanoparticle has spin=1/2; the free spin in 
the nanoparticle can form a singlet with the electron on 
the island.  However, for $N$=odd the nanoparticle has a 
singlet ground state; the coupling between the nanoparticle 
and the island is through virtual fluctuations.  The 
differential capacitance vs. gate voltage is plotted 
in Fig.~\ref{fig:capacitance}

\vspace{.1in}
\vspace{.1in}

%%%%%%%%%%%%%%%%%%%%%%%%%%%%%%%%%%%%%%%%%%%%%%%%%%%%%%%%%%%%%%%%%%%%%%%%%%%
%%%%%%%%%%%%%%%%%%%%%%%%%%%%%%%%%%%%%%%%%%%%%%%%%%%%%%%%%%%%%%%%%%%%%%%%%%%
  
In conclusion, carbon nanotube SETs\cite{postma} may be promising 
for studying the Kondo effect.  
With semi-infinite leads, this system allows for a detailed 
investigation of the 2-channel Kondo fixed point.
We also considered the Kondo effect in a finite-sized nanotube --- 
a Kondo box.  Here, we saw that the results depend on whether the 
total number of particles is even or odd.  
Finally, it is worth noting that generalizations of this device
could allow for the study of other related phenomena.  For example,
introducing two islands in the nanotube could allow one to study
two-impurity Kondo physics, or more generally, the properties
of coupled quantum dots.  

%%%%%%%%%%%%%%%%%%%%%%%%%%%%%%%%%%%%%%%%%%%%%%%%%%%%%%%%%%%%%%%%%%%%%%%%%%%
%%%%%%%%%%%%%%%%%%%%%%%%%%%%%%%%%%%%%%%%%%%%%%%%%%%%%%%%%%%%%%%%%%%%%%%%%%%

% \section*{Acknowledgements}

EHK is grateful to H. Paik for bringing Ref.~\ref{postma} to 
his attention.
This work was supported by the NSERC of Canada (EHK and CK), and 
the Spanish grant PB98-0685 (GS).
 
%%%%%%%%%%%%%%%%%%%%%%%%%%%%%%%%%%%%%%%%%%%%%%%%%%%%%%%%%%%%%%%%%%%%%%%%%%%
%%%%%%%%%%%%%%%%%%%%%%%%%%%%%%%%%%%%%%%%%%%%%%%%%%%%%%%%%%%%%%%%%%%%%%%%%%%

%%%%%%%%%%%%%%%%%%%%%%%%%%%%%%%%%%%%%%%%%%%%%%%%%%%%%%%%%%%%%%%%%%%%%%%%%%%
%%%%%%%%%%%%%%%%%%%%%%%%%%%%%%%%%%%%%%%%%%%%%%%%%%%%%%%%%%%%%%%%%%%%%%%%%%%

\end{multicols}
\end{document}